\documentclass[aps,pre,twocolumn,showpacs]{revtex4}
\usepackage{amssymb}
\usepackage{eurosym}
\usepackage{epsfig}

\begin{document}

\title{Expansion of the strongly interacting superfluid Fermi gas:
symmetries and self-similar regimes}
\author{E.A. Kuznetsov$^{(a),(b),(c)}$, M.Yu. Kagan $^{(d),(e),(f)}$ and
A.V. Turlapov $^{(e)}$}
\affiliation{$^{(a)}$ P.N. Lebedev Physical Institute RAS, Moscow, Russia\\
$^{(b)}$ L.D. Landau Institute for Theoretical Physics RAS, Chernogolovka,
Moscow region, Russia\\
$^{(c)}$ Skolkovo Institute of Science and Technology, Moscow, Russia\\
$^{(d)}$ P.L. Kapitza Institute of Physical Problems RAS, Moscow, Russia\\
$^{(e)}$ Institute of Applied Physics RAS, Nizhniy Novgorod, Russia \\
$^{(f)}$ National Research University ''Higher School of Economics'',
Moscow, Russia}

\begin{abstract}
We consider an expansion of the strongly interacting superfluid Fermi gas in
a vacuum, assuming absence of the trapping potential, in the so-called
unitary regime (see, for instance, \cite{pitaevskii2008superfluid}) when the
chemical potential $\mu \propto \hbar^2n^{2/3}/m$ where $n$ is the density
of the Bose-Einstein condensate of Cooper pairs of fermionic atoms. In low
temperatures, $T\to 0$, such expansion can be described in the framework of
the Gross-Pitaevskii equation (GPE). Because of the chemical potential
dependence on the density, $\sim n^{2/3}$, the GPE has additional
symmetries, resulting in the existence of the virial theorem \cite%
{vlasov1971averaged}, connecting the mean size of the gas cloud and its
Hamiltonian. It leads asymptotically at $t\to\infty$ to the gas cloud
expansion, linearly growing in time. We study such asymptotics, and reveal
the perfect match between the quasi-classical self-similar solution and the
asymptotic expansion of the non-interacting gas. This match is governed by
the virial theorem, derived through utilizing the Talanov transformation 
\cite{talanov1970focusing}, which was first obtained for the stationary
self-focusing of light in media with a cubic nonlinearity due to the Kerr
effect. In the quasi-classical limit, the equations of motion coincide with
3D hydrodynamics for the perfect monoatomic gas with $\gamma=5/3$. Their
self-similar solution describes, on the background of the gas expansion, the
angular deformities of the gas shape in the framework of the
Ermakov--Ray--Reid type system.
\end{abstract}

\maketitle


\section{Introduction}

Since the discovery of Bose-Einstein condensation (BEC) in alkali gases of
bosonic isotopes $^{7}Li,^{23}Na,^{87}Rb$ \cite{anderson1995observation,
bradley1995evidence, davis1995kb} the time of flight experiments connected
with the expansion of the Bose condensate cloud from the trap, when the
trapping potential is switched off, served as one of the important proves of
the superfluid transition in the gas. Note that in these experiments the
velocity distribution in the gas expansion has a typical bimodal shape which
corresponds to two components - a normal and a superfluid ones. The velocity
distribution of the normal component has a thermal (Maxwell-type) shape
while the velocity of the superfluid component is governed only by the
interaction parameter of the weakly non-ideal (in the Gross-Pitaevskii
meaning) Bose gas and the total number of particles.

The first scaling time-dependent solutions both for the condensate and
thermal gas in the hydrodynamic regime for the anisotropic trap was obtained
by Yu.Kagan, Surkov and Shlyapnikov \cite{kagan1997evolution}. In
particular, in \cite{kagan1997evolution} the spectrum of breathing modes of
the oscillating type in the trapping potential was also determined. Later on
the self-similar regimes were observed in the experiments of the Thomas'
group \cite{o2002observation} for the anisotropic expansion of strongly
interacting degenerate Fermi gas of atoms $^{6}Li$ from the optical trap.
The measurements of this group were performed in the regime of Feshbach
resonance \cite{feshbach1958unified,chin2010feshbach} which corresponds to
the BCS-BEC crossover \cite{nozieres1985bose, leggett1980cooper,
combescot2006self} between extended Cooper pairs of fermionic atoms (BCS)
and BEC of tightly bound fermion pairs (molecules or dimers $^{6}Li_{2}$).

Note that while exploiting Feshbach resonance it is possible to reach large
absolute values of the $s$-wave scattering length $|a_s|$ and thus to
increase sharply the critical temperature of the superfluid transition for
the fixed particle density $n$ in the trap reaching experimentally
accessible values of critical temperature $T_{c}$. Remind that BCS phase of
extended pairs on the phase diagram of the BCS-BEC crossover corresponds to
the positive values of the chemical potential $\mu >0$ and negative values
of the scattering length $a_s<0$. In the same time the BEC phase of local
pairs vice versa corresponds to a positive scattering length $a_s>0$ and
negative chemical potential $\mu <0$. For small positive values of the gas
parameter $0<a_sk_{F}\ll 1$ (where $p_{F}=\hbar k_{F}$ is Fermi momentum) we
are in the dilute BEC domain which describes a Bogoliubov gas of weakly
repulsive composed bosons (local fermion pairs). In the same time, small
negative values of the gas parameter $-1\ll (a_sk_{F})^{-1}<0$ corresponds
to the dilute BCS regime describing weakly attractive degenerate Fermi gas
of atoms.

\begin{figure}[t]
\centering
\includegraphics[width=6.7cm]{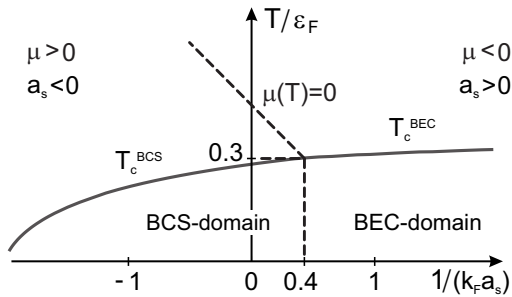}
\caption{\textit{(Color on-line)} The phase diagram of the BCS-BEC crossover
in the $T$- matrix approximation for the 3D interacting Fermi gas in the
regime of Feshbach resonance \protect\cite{combescot2006self,
kagan2019crossover}.}
\label{fig:fig1}
\end{figure}

In the regime of Feshbach resonance it is convenient to consider the phase
diagram of the BCS-BEC crossover in terms of the dimensionless temperature $%
T/\varepsilon _{F}$ (the vertical axis in Fig.~\ref{fig:fig1}) and the
inverse gas parameter $(a_sk_{F})^{-1}$ which scales linearly with the shift
of external magnetic field $\Delta B=B-B_{0}$ from the resonance value $%
B_{0} $ at the horizontal axis (see \cite{combescot2006self}). Then on the
phase diagram between dilute BCS and BEC domains there appears an
intermediate region of strong correlations where the inverse gas parameter
varies in the interval $-1<(ak_{F})^{-1}<1$ . In this region it is difficult
to develop rigorous diagrammatic expansions and only some reasonable
(conserving) approximations like the self-consistent $T$-matrix
approximations \cite{combescot2006self} are available. 
However, we have one special point in this regime which lies in the middle
of strongly interacting region and corresponds to the unitarian limit $%
(a_sk_{F})^{-1}\rightarrow 0$ where exact (and universal) results can be
obtained (see e.g. \cite{pitaevskii2008superfluid} for the review). The
reason for the universality of the unitarian limit is connected with the
fact that there is no other energy scales besides Fermi energy $\varepsilon
_{F}$ in this point. Thus, both the chemical potential $\mu $ and the
critical temperature $T_{c}$ in the unitarian limit scale linearly with $%
\varepsilon _{F}$.

Note also that at the middle of 1990-es in the field of ultra-cold quantum
gases only anisotropic traps were experimentally available. Anisotropy
degree for the gas clouds distributions in traps changed from 3D (with weak
anisotropy) up to quasi-1D (with strong anisotropy for cigar-shaped traps).
For instance, the cigar-shaped traps were considered in the Thomas' group
experiments \cite{o2002observation}. Almost spherical trap, to our
knowledge, appeared only in 2015 (see \cite{lobser2015observation}). For the
spherical traps all three trapping frequencies are equal, $\omega
_{x}=\omega _{y}=\omega _{z}$. For the cigar-shaped traps we have the
following hierarchy of the trapping frequencies: $\omega _{z}\ll \{\omega
_{x},\omega _{y}\}$.

Quite recently also the disk-shaped traps with quasi-2D cloud distributions
become experimentally available (see \cite{turlapov2017fermi} and references
therein). Parameters of these traps correspond to inverse hierarchy of the
trapping frequencies with respect to cigar-shaped traps, namely $\omega
_{z}\gg \{\omega _{x},\omega _{y}\}$. In this paper, we are discussing also
expansion of the superfluid Fermi gas into vacuum from disk-shaped traps 
\cite{turlapov2018expansion}. It is interesting to note that the simplest
ballistic picture of the Fermi gas expansion for the disk-shaped superfluid
in the unitarian regime considered in \cite{turlapov2018expansion} coincides
with the exact quasi-classical self-similar solution first found by Anisimov
and Lysikov for the hydrodynamic expansion of the classical gas with
adiabatic constant $\gamma =5/3$ \cite{anisimov1970expansion} (in the
absence of vorticity).

In this paper, we consider an expansion of the strongly interacting
superfluid Fermi gas in a vacuum in the unitary regime when the chemical
potential $\mu \propto \hbar ^{2}n^{2/3}/m$ where $n$ is the density of the
Bose-Einstein condensate of Cooper pairs of fermionic atoms assuming
temperature $T\to 0$. Such expansion can be described in the framework of
the Gross-Pitaevskii equation (GPE). Because of the chemical potential
dependence on the density $\sim n^{2/3}$ the GPE has additional symmetries
resulting in existence of the virial theorem \cite{vlasov1971averaged}
connected the mean size of the gas cloud and its Hamiltonian. It leads
asymptotically at $t\rightarrow \infty $ to the linear in time expansion of
the gas. We carefully study such asymptotics and reveal a perfect matching
between the quasi-classical self-similar solution and the asymptotic
expansion of the non-interacting gas.

The paper is organized as follows. In the next Section we discuss the
problem concerning symmetries of the GPE in the unitarian limit and how
these symmetries are connected with those found for the nonlinear
Schrodinger equation (NLSE) in the critical case when the virial theorem can
be applied for description of the stationary self-focusing of light in media
with the Kerr nonlinearity \cite{vlasov1971averaged} and symmetries in the
case of ideal monoatomic gas. Section 3 mainly deals with self-similar
solution of the anisotropic type of the quasi-classical GPE in the unitarian
limit for expansion of the Fermi superfluid gas. This solution describes the
angular deformations of the gas shape on the background of the gas
expansion. In Section 4, we discuss in which extent the analytical results
obtained in the previous sections are related with experimental data.
Conclusion summarizes all results of this paper.

\section{Symmetries and integrals of motion}

First of all, we would like to remind that the topic of gas expansion was
very popular in the hydrodynamic content in 60-s of the XXth century. The
first classical works were performed by L.V. Ovsyannikov (1956) \cite%
{ovsyannikov1956solution} and F.J. Dyson (1968) \cite{dyson1968dynamics}.
These studies had a lot of applications not only in hydrodynamics but also
in astrophysics (see, e.g. the original paper by Ya. B. Zel'dovich \cite%
{zel1964newtonian}).

In 1970 S.I. Anisimov and Yu.I. Lysikov \cite{anisimov1970expansion}
discovered very interesting phenomenon connected with the nonlinear angular
deformation of the gas cloud while its expansion. Such behavior directly
follows from their remarkable solution for a gas with specific heat ratio $%
5/3 $ (see e.g. reviews \cite{bogoyavlensky1979nonlinear,
borisov2008hamiltonian} and references therein). This result, as was pointed
out by I.E. Dzyaloshinskii (private communication, 1970), represents a
consequence of the symmetry which is well known in quantum mechanics for
motion of a non-relativistic particle in the potential $V(r)=\beta /r^{2}$.
This symmetry, independent on the sign of $\beta $, is dilatations of both
spatial coordinates and time for which $\mathbf{r}\rightarrow \alpha \mathbf{%
r}$ and $t\rightarrow \alpha ^{2}t$ where $\alpha $ is a scaling parameter.
Indeed, such symmetry first time was exploited by V.P. Ermakov in 1880 \cite%
{Ermakov1880differential} to construct solutions for some mechanical systems
including motion of a particle in the potential which is a combination of
the oscillator potential and $V(r)=\beta /r^{2}$. In seventies of the XXth
century, Ray and Reid \cite{ray1979more} rediscovered the Ermakov results.
Now all such equations are accepted to call the Ermakov-Ray-Reid systems
(see, e.g. \cite{rogers1996multi} and references therein). As we will show
in this paper, this additional symmetry for the GPE takes place for both
attractive and repulsive interactions (in optical content, corresponding to
focusing and defocusing nonlinearities). Note that, in quantum mechanics
(see \cite{landau1965quantum}), for the attractive potential, $V<0$, with
constant $|\beta |$ larger some critical value ($=\hbar ^{2}/\left(
8m\right) $) the quantum falling of a particle with mass $m$ into the center
is possible which can be understood as collapse. Moreover, this falling
becomes more quasiclassical while approach the center. In the case of the
Gross-Pitaevskii equation (GPE) \cite{gross1961structure,
pitaevskii1961vortex} which can be applied for description of nonlinear
dynamics of the Bose condensate for diluted gases, the kinetic energy has
the same scaling as in the usual quantum mechanics, i.e. $\propto \alpha
^{-2}$. The nonlinear interaction term in the GPE, due to the $s$%
-scattering, has a scaling $\propto \alpha ^{-d}$ which appears from the
conservation of the total number of particles $N=\int |\psi |^{2}d\mathbf{r}$
with $d$ being the space dimension and $\psi $ the wave function of the Bose
condensate. Thus, at $d=2$ only we have the situation analogous to that in
the quantum mechanics for potentials $V(r)=\beta /r^{2}$. This case, as it
was first time demonstrated by Vlasov, Petrishchev and Talanov \cite%
{vlasov1971averaged} for the 2D nonlinear Schrodinger equation, is very
special for which the so-called virial relation is valid: 
\begin{equation}
m\frac{d^{2}}{dt^{2}}\int r^{2}|\psi |^{2}d\mathbf{r}=4H,  \label{eq.1}
\end{equation}%
where the Hamiltonian $H$ in the case of the GPE for the Bose condensate has
the form 
\[
H=\int \left[ \frac{\hbar ^{2}}{2m}|\nabla \psi |^{2}+g|\psi |^{4}\right] d%
\mathbf{r}. 
\]%
Here the coupling coefficient $g=4\pi \hbar ^{2}a_{s}/m$ with $a_{s}$ being
scattering length and $m$ particle mass. It is necessary to emphasize that
the virial relation (\ref{eq.1}) is valid for any sign of $g$. The only
restriction follows from the requirement of convergence of the integrals
standing in (\ref{eq.1}). It is worth noting that in classical mechanics the
virial theorem establishes the ratio between mean values of the total
kinetic and potential energies. The simplest way to derive this theorem is
calculation of the second time derivative of a moment of inertia (this
results in the virial relation like Eq. (\ref{eq.1})) and then averaging it
in time. Further we will call relation (\ref{eq.1}) as the virial theorem.

In this paper we consider another example of the same symmetry, when the
generalized Gross-Pitaevskii equation \cite{pitaevskii2008superfluid} can be
applied for description of the strongly interacting Fermi gas in the
superfluid phase (at $T=0$): 
\begin{equation}
i\hbar \frac{\partial \psi }{\partial t}=-\frac{\hbar ^{2}}{2(2m)}\Delta
\psi +\mu (n)\psi ,  \label{eq.2}
\end{equation}%
where $\psi $ is the wave function of the Bose condensate of fermion pairs, $%
m$ is a fermion mass ($2m$ is a mass of a fermion pair), $\mu $ is the
chemical potential. In the unitarian limit (when $(k_{F}a)^{-1}\rightarrow 0$%
) the chemical potential reads (see e.g. \cite{pitaevskii2008superfluid}):%
\begin{equation}
\mu (n)=2(1+\beta )\varepsilon _{F},  \label{eq.3}
\end{equation}%
where universal interaction parameter $\beta =-0.63$, in accordance with 
\cite{JosephSound}, \cite{Grimmbeta,ZwierleinLambda2012}, \cite%
{JochimNewLiFeshbach2013}, and local Fermi energy%
\[
\varepsilon _{F}=\frac{\hbar ^{2}}{2m}\left( 6\pi ^{2}n\right) ^{2/3}.
\]
Here $n=|\psi |^{2}$ is concentration of fermionic pairs. Below we will
normalize density $n$ by its initial maximum value $n_{0}$, inverse time $%
t^{-1}$ by $\frac{\hbar }{2m}n_{0}^{2/3}$ and coordinate $r$ by $%
n_{0}^{-1/3} $. In these new (dimensionless) units equation (\ref{eq.2})
reads as 
\begin{equation}
i\frac{\partial \psi }{\partial t}=-\frac{1}{2}\Delta \psi +\mu (n)\psi
\label{GP-1}
\end{equation}%
where 
\begin{equation}
\mu (n)=2(1+\beta )\left( 6\pi ^{2}n\right) ^{2/3}.  \label{chemical}
\end{equation}

Choosing the standard ansatz for the $\psi $-function, $\psi =\sqrt{n(r,t)}%
\exp \left( i\varphi (r,t)\right) $, and separating then real and imaginary
parts in (\ref{eq.2}) we get the system of continuity and Euler (eiconal)
equations 
\begin{eqnarray}
\frac{\partial n}{\partial t}+\left( \nabla \cdot n\nabla \varphi \right)
&=&0,  \label{eq.5} \\
\frac{\partial \varphi }{\partial t}+\left[ \mu (n)+\frac{\left( \nabla
\varphi \right) ^{2}}{2}+T_{QP}\right] &=&0,  \label{eq.6}
\end{eqnarray}%
where $\mathbf{v=}\nabla \varphi $ has a meaning of velocity. Here we used
the condition of the absence of vortices $\nabla \times \mathbf{v}=0$.

Note that the term in Eq. (\ref{eq.6}) represents the quantum pressure given
by%
\begin{equation}
T_{QP}=-\frac{\Delta \sqrt{n}}{2\sqrt{n}}.  \label{eq.7}
\end{equation}%
Throughout the main part of the present paper we will neglect this term and
will discuss its possible role in two last Sections. Neglecting quantum
pressure corresponds to the quasi-classical (or eikonal ) approximation
(called also the time-dependent Thomas-Fermi approximation) which assumes
more rapid space and time variations of phase (larger phase gradients and
time-derivatives) in comparison with the space and time variations of the
modulus of the $\psi $-function in Eq.(\ref{GP-1}).

It is important to emphasize that the generalized Gross-Pitaevskii equation (%
\ref{GP-1}) coincides with the nonlinear Schroedinger equation (NLSE) widely
used in nonlinear optics and plasma physics. It is convenient to exclude in (%
\ref{chemical}) the factor $2(1+\beta )\left( 6\pi ^{2}\right) ^{2/3}$ by
simple rescaling of the density $n$, 
\[
2(1+\beta )\left( 6\pi ^{2}n\right) ^{2/3}\rightarrow \frac{5}{3}n^{2/3} 
\]%
so that equation (\ref{GP-1}) takes the standard form accepted for the NLSE, 
\begin{equation}
i\frac{\partial \psi }{\partial t}+\frac{1}{2}\Delta \psi -(\upsilon
+1)|\psi |^{2\upsilon }\psi =0,  \label{eq.8}
\end{equation}%
with the exponent $\upsilon =2/3$. This equation can be written in the
Hamiltonian form%
\[
i\frac{\partial \psi }{\partial t}=\frac{\delta H}{\delta \psi ^{\ast }}, 
\]%
where Hamiltonian%
\begin{equation}
H=\int \left[ \frac{1}{2}|\nabla \psi |^{2}+|\psi |^{2(\upsilon +1)}\right] d%
\mathbf{r,}  \label{Ham-psi}
\end{equation}%
with the first term coinciding with the total kinetic energy and the second
one responsible for nonlinear interaction of the repulsion type. After
applying the transformation $\psi =\sqrt{n(r,t)}\exp \left( i\varphi
(r,t)\right) $ equations for density $n$ and phase $\varphi $ remain the
Hamiltonian form,%
\begin{equation}
\frac{\partial n}{\partial t}=\frac{\delta H}{\delta \varphi },\,\,\frac{%
\partial \varphi }{\partial t}=-\frac{\delta H}{\delta n}  \label{Ham-eqs}
\end{equation}%
where the Hamiltonian coincides with (\ref{Ham-psi}). In terms of $n$ and $%
\varphi $, $H$ takes the form 
\[
H=\int \left[ \frac{n\left( \nabla \varphi \right) ^{2}}{2}+\frac{\left(
\nabla \sqrt{n}\right) ^{2}}{2}+n^{\upsilon +1}\right] d\mathbf{r.} 
\]%
The Hamiltonian equations of motion (\ref{Ham-eqs}) are the same Eqs. (\ref%
{eq.5}) and (\ref{eq.6}) transformed under simple rescaling; $n$ and $%
\varphi $ in this case play the role of canonically conjugated quantities.

The second term in $H$ is responsible for the quantum pressure in Eq. (\ref%
{eq.6}). In the quasiclassical limit (the Thomas-Fermi approximation), this
term becomes small and can be neglected so that we arrive at the
hydrodynamic equations for potential flow of monoatomic gas with specific
heat ratio (adiabatic index) $\gamma =5/3$ ( $\nu =2/3$). This $\gamma$ is
remarkable for both NLSE and its quasiclassical limit. It turns out that the
equations of motion in this case have two additional symmetries. The first
symmetry forms dilatation group of the scaling type: $\mathbf{r\rightarrow }%
\alpha \mathbf{r}$ and $t\mathbf{\rightarrow }\alpha ^{2}t$ . In usual
quantum mechanics, this symmetry appears for the potential $V(r)\sim r^{-2}$
independently on both the potential sign (attraction or repulsion) and space
dimension $d$. However, for the NLSE (\ref{eq.8}) such symmetry appears as a
result of the conservation of the total number of particles $N=\int |\psi
|^{2}d\mathbf{r}$ so that at $d=3$ only the nonlinear potential $\sim |\psi
|^{4/3}$ in Eq. (\ref{eq.8}) has the same scaling as the Laplace operator $%
\Delta $. At $d=2$ such symmetry takes place already for the nonlinear
potential $\sim |\psi |^{2}$ (in this case the NLSE describes the stationary
self-focusing of light in a media with the Kerr nonlinearity). In the
general case the dilatation symmetry arises at $\nu =2/d$ (see, for
instance, \cite{kuznetsov1985talanov, rasmussen1986blow}). The second symmetry of the conformal type first time was
found by V.I. Talanov for the cubic NLSE at $d=2$ \cite{talanov1970focusing}
(1970) and called now the Talanov transformations. In optical content these
are the lens transformations well known in a linear optics.

These symmetries are of the Noether type and generate two additional
integrals of motion. They can be obtained from the virial theorem (\ref{eq.1}%
) (first time obtained for the 2D cubic NLSE in \cite{vlasov1971averaged}),
after twice integration in time (dimensionless variables): 
\begin{equation}
\int r^{2}|\psi |^{2}d\mathbf{r=}2Ht^{2}+C_{1}t+C_{2}.  \label{eq.13}
\end{equation}%
Hence we get asymptotically at $t\rightarrow \infty $, independently on $%
C_{1}$ and $C_{2}$, 
\[
\int r^{2}|\psi |^{2}d\mathbf{r\rightarrow }2Ht^{2}.  
\]
Therefore the mean size (indeed, r.m.s) of the gas cloud varies at large $t$
linearly in time, 
\begin{equation}
\left\langle r^{2}\right\rangle ^{1/2}\mathbf{\propto }t\sqrt{2H/N}.
\label{eq.15}
\end{equation}%
This result is very important also since it perfectly matches
quasi-classical solution in Eq.(\ref{eq.15}) with the linearly varying
solutions for the non-interacting particles (ballistic expansion).

It should be emphasized that the virial theorem (\ref{eq.1}) is the exact
result, it can be applied in particular in the quasi-classical limit also
when the quantum pressure term in $H$ is eliminated. The latter corresponds
to the classical gas expansion with $\gamma =5/3$. In this case Anisimov and
Lysikov (1970) \cite{anisimov1970expansion} constructed exact anisotropic
self-similar solution based, in fact, on existence of two integrals of
motion $C_{1}$ and $C_{2}$ (see the next subsection). This solution
describes the gas expansion in time in correspondence with (\ref{eq.13})
with nonlinear angular deformation of the gas shape.

\section{Self-similar quasi-classical solution}

Note that the corresponding system of gas dynamics equations in the
quasi-classical limit is described by Eqs. (\ref{eq.5},\ref{eq.6},\ref{eq.8},%
\ref{Ham-psi}) by neglecting quantum pressure in (\ref{eq.6}) and with the
eikonal equation for the phase which in the unitarian limit (for $\upsilon
+1=5/3$) is given by%
\begin{equation}
\frac{\partial \varphi }{\partial t}+\frac{\left( \nabla \varphi \right) ^{2}%
}{2}+\frac{5}{3}n^{2/3}=0.  \label{eq.16}
\end{equation}
Let us search for a solution of these equations in the self-similar form
(see e.g. \cite{anisimov1970expansion, zakharov1986quasi,  rypdal1986blow, menotti2002c}),%
\begin{equation}
n=\frac{1}{a_{x}a_{y}a_{z}}f\left( \frac{x}{a_{x}},\frac{y}{a_{y}},\frac{z}{%
a_{z}}\right) ,  \label{eq.17}
\end{equation}%
assuming that three scaling parameters $a_{x},a_{y},a_{z}$ are functions of
time. Note that the ansatz (\ref{eq.17}) conserves the total number of
particles.

Calculating the (partial) time derivative from density we get%
\begin{equation}
\frac{\partial n}{\partial t}=-\frac{1}{a_{x}a_{y}a_{z}}\sum_{i}\frac{\dot{%
a_{i}}}{a_{i}}\frac{\partial }{\partial \xi _{i}}(f\xi _{i}),  \label{eq.18}
\end{equation}%
where $i=(x,y,z)$. Here we introduced convenient notations for the
self-similar variables%
\[
\xi _{x}=\frac{x}{a_{x}},\,\,\xi _{y}=\frac{y}{a_{y}},\,\,\xi _{z}=\frac{z}{%
a_{z}}. 
\]%
Then the continuity equation admits integration resulting in relations for
the phase $\varphi $ 
\[
\varphi =\varphi _{0}(t)+\sum_{l}\frac{\dot{a_{l}}a_{l}}{2}\xi _{l}^{2}, 
\]%
where function $\varphi _{0}(t)$ can be found after substitution in the
eikonal equation. Calculating first two terms in (\ref{eq.16}) yields:%
\[
\frac{\partial \varphi }{\partial t}+\frac{\left( \nabla \varphi \right) ^{2}%
}{2}=\frac{d\varphi _{0}(t)}{dt}+\sum_{l}\frac{\ddot{a_{l}}a_{l}}{2}\xi
_{l}^{2}. 
\]%
The third term in (\ref{eq.16}) reads%
\[
\frac{5}{3}n^{2/3}=\frac{5}{3}\frac{1}{\left( a_{x}a_{y}a_{z}\right) ^{2/3}}%
\left[ f\left( \xi _{x},\xi _{y},\xi _{z}\right) \right] ^{2/3}. 
\]%
Hence in order to satisfy the self-similar ansatz one needs to require that%
\begin{eqnarray*}
&&\frac{d\varphi _{0}(t)}{dt}=-\frac{5}{3}\frac{1}{\left(
a_{x}a_{y}a_{z}\right) ^{2/3}}f(0)^{2/3}, \\
&&\sum_{l}\frac{\ddot{a_{l}}a_{l}}{2}\xi _{l}^{2}=\frac{5}{3\left(
a_{x}a_{y}a_{z}\right) ^{2/3}}[f\left( 0\right) ^{2/3}-f\left( \xi \right)
^{2/3}].
\end{eqnarray*}%
Hence we conclude that 
\begin{equation}
\ddot{a_{x}}a_{x}=\ddot{a_{y}}a_{y}=\ddot{a_{z}}a_{z}=\frac{\lambda }{\left(
a_{x}a_{y}a_{z}\right) ^{2/3}}  \label{eq.27}
\end{equation}%
where $\lambda $ is arbitrary positive constant. For $f\left( \mathbf{\xi }%
\right) $ we have%
\begin{equation}
f\left( \mathbf{\xi }\right) =\left[ f\left( 0\right) ^{2/3}-\frac{3\lambda 
}{10}\mathbf{\xi }^{2}\right] ^{3/2},  \label{function-f}
\end{equation}%
where further we will put $f\left( 0\right) =1$. Respectively, the density
is written as 
\begin{equation}
n=\frac{1}{a_{x}a_{y}a_{z}}\left[ 1-\frac{3\lambda }{10}\mathbf{\xi }^{2}%
\right] ^{3/2}  \label{density}
\end{equation}%
here the constant $\lambda $ will be found from the initial condition.

We will assume that initially the density distribution is defined from the
Thomas-Fermi approximation. In the presence of harmonic trap, at the
stationary state we have the equilibrium condition%
\[
\mu (n)=\mu (n_{0})-m\sum \omega _{i}^{2}x_{i}^{2}, 
\]%
where $\mu (n)$ is defined by (\ref{chemical}). Remind that because of
pairing in this expression $2m$ stands instead of $m$. This gives the
initial density distribution:%
\[
n=n_{0}\left[ 1-\frac{m\omega _{m}^{2}n_{0}^{-2/3}}{\mu (n_{0})}\sum \xi
_{i}^{2}\right] ^{3/2}, 
\]%
where $\omega _{m}=\max (\omega _{i})$, $a_{i}(0)=\omega _{m}/\omega _{i}$.
This profile matches precisely with self-similar solution (\ref{density}) at 
$t=0$. Hence we have that 
\[
\lambda =\frac{10m\omega _{m}^{2}n_{0}^{-2/3}}{3\mu (n_{0})} 
\]%
or in terms of $N$ 
\[
\lambda =\frac{5}{6}\left( \frac{\pi ^{2}}{N}\right) ^{2/3}\left( \frac{%
\omega _{m}^{3}}{\omega _{x}\omega _{y}\omega _{z}}\right) ^{2/3}. 
\]
Function $f\left( \mathbf{\xi }\right) $ (\ref{function-f}) is spherically
symmetric with respect to $\mathbf{\xi }$ . It varies from $1$ at $\xi =0$
up to zero at $\xi _{\max }=$\ $\sqrt{10/3\lambda }$; above $\xi _{\max }$
the density $n$ is equal to zero (see Fig.2). 
\begin{figure}[b]
\label{fig2} 
\centerline{
		\includegraphics[width=0.45\textwidth]{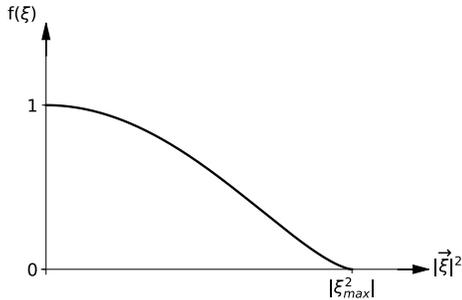}}
\caption{The behavior of the density factor $f(\protect\xi )$ (arbitrary
units).}
\end{figure}
In accordance with (\ref{eq.27}) dynamics of three scaling parameters $%
a_{i}(t)$ ($i=1,2,3$) is described by the Newton equations for motion of a
particle 
\begin{equation}
{\ddot{a_{i}}=-\frac{\partial U}{\partial a_{i}}},  \label{eq.30}
\end{equation}%
where potential 
\begin{equation}
U=\frac{3\lambda }{2\left( a_{x}a_{y}a_{z}\right) ^{2/3}}.  \label{eq.31}
\end{equation}%
It is worth noting that at the point $\xi =\xi _{\max }$ the obtained
quasiclassical solution given by (\ref{eq.17}-\ref{eq.30}) breaks down that
follows from estimation of the quantum pressure term which becomes
infinitely large. In this case, $\xi =\xi _{\max }$ plays the role of a
reflection point in the usual quasiclassical approximation in quantum
mechanics. This means that at the vicinity $\Delta \xi $ around $\xi =\xi
_{\max }$ one needs to match the constructed solution at $\xi <\xi _{\max }$
(inner region) with that at $\xi >\xi _{\max }$ (outer region) where we
should neglect nonlinearity in the NLSE (free Schrodinger equation). This
problem was discussed in details in \cite{zakharov1986quasi} for the strong
collapse regime in the supercritical NLSE with $d=3$ and $\nu =1$. In the
given case, the matching problem can be also resolved if $\Delta \xi \ll \xi
_{\max }$ when the matching solution is expressed in terms of the Painleve
function. It should be noted that for the NLSE (\ref{eq.8}) with account of $%
T_{QP}$ the scaling remains the same as for the quasiclassical solution and
by this reason $\Delta \xi $ can be considered as time-independent quantity
so that the ratio between $\Delta \xi $ and $\xi _{\max }$ remains in time
the same, unlike \cite{zakharov1986quasi} where this ratio in 3D collapse
regime is time-dependent and vanishes while approaching the collapse time.

We should notice also that near $\xi =\xi _{\max }$ the unitarian limit is
not also applicable because $k_{F}$ becomes infinitely large. Near this
point, however, the nonlinear term is small and we again return to the same
matching problem like in the previous case.

\subsection{The virial theorem for the scaling parameters}

It is more or less evident that Newton equations (\ref{eq.30}) have to have
the same symmetry properties as the original NLSE (\ref{eq.18}) what can be
easily verified. Note first that for Eqs. (\ref{eq.30}) the energy integral
is written in the standard form 
\[
E=\frac{1}{2}\sum_{i=1,2,3}\dot{a_{i}}^{2}+\frac{3\lambda }{2\left(
a_{x}a_{y}a_{z}\right) ^{2/3}}. 
\]%
Secondly, for Eqs. (\ref{eq.30}) by direct calculation it is possible to get
the virial theorem (\ref{eq.1}), written in terms of $a_{i}$. For $\sum
a_{i}^{2}$ we have%
\[
\frac{d^{2}}{dt^{2}}\sum_{i}a_{i}^{2}=2\sum_{i}\left[ \left( \frac{da_{i}}{dt%
}\right) ^{2}+a_{i}\frac{d^{2}a_{i}}{dt^{2}}\right] 
\]%
Then substitution of (\ref{eq.30}) into this relation gives finally 
\[
\frac{d^{2}}{dt^{2}}\sum_{i}a_{i}^{2}=2\sum_{i}\left( \frac{da_{i}}{dt}%
\right) ^{2}+\frac{6\lambda }{\left( a_{x}a_{y}a_{z}\right) ^{2/3}}=4E, 
\]%
that coincides with the virial identity (\ref{eq.1}). Its twice integration
gives two constants $C_{1}$ and $C_{2}$ (integrals of motion): 
\begin{equation}
\sum_{i}a_{i}^{2}=2Et^{2}+C_{1}t+C_{2}.  \label{quasi-virial}
\end{equation}%
Hence%
\begin{eqnarray}
C_{1} &=&\frac{d}{dt}\sum_{i}a_{i}^{2}-4Et,  \label{C-1and C-2} \\
C_{2} &=&\sum_{i}a_{i}^{2}-2Et^{2}-C_{1}t.
\end{eqnarray}%
In the isotropic (spherically symmetric) case when $a_{x}=a_{y}=a_{z}\equiv
a $ the equations of motion transforms into one equation 
\begin{equation}
\ddot{a}=\frac{\lambda }{a^{3}}  \label{eq.38}
\end{equation}%
with the energy $E=\frac{3}{2}\left( \dot{a}^{2}+\frac{\lambda }{a^{2}}%
\right) $ and $3a^{2}=2Et^{2}+C_{1}t+C_{2}$. From the second relation we
immediately have that gas cloud expands in radial direction asymptotically
at $t\rightarrow \infty $ with constant velocity 
\begin{equation}
v_{\infty }=\sqrt{2E/3}  \label{eq.41}
\end{equation}%
(ballistic regime). This result is in agreement with the virial theorem (\ref%
{eq.15}).

If we change the sign of the potential in Eq.(\ref{eq.38}) then we get the
falling of the particle on the potential center which, as known in quantum
mechanics, becomes more quasiclassical while approaching the center (see 
\cite{landau1965quantum}).

For the expansion of a noninteracting gas from a harmonic potential 
\[
\sqrt{\left\langle x_{i}^{2}\right\rangle }\propto \left( \sqrt{\hbar \omega
_{i}/2m}\right) t 
\]%
and $v_{\infty }=const$ in agreement with our intuitive considerations and
with Eq.(\ref{eq.41}) as well. (Let us remind that for quasi-2D disk-shaped
traps the trapping frequency $\omega _{z}\gg \omega _{x}\simeq \omega _{y}$%
). Thus, we have almost perfect matching of ballistic results for
non-interacting gas and quasi-classical results derived for
strongly-interacting Fermi gas in the eikonal approximation.

It is worth noting that in the virial relation (\ref{quasi-virial}) besides
total energy $E$ there enter two more constants (integrals of motion) $%
C_{1},C_{2}$ . In principle, if $C_{1}>0$ then the solution with $%
a^{2}\propto C_{1}t$ is possible for some intermediate times, but not
initially. The regime with $a\propto \left( t_{0}-t\right) ^{1/2}$ is
typical for weak self-similar collapse (see \cite{zakharov1986quasi,
zakharov2012solitons}).

\subsection{Anisotropic self-similar solution}

The simplest anisotropic case corresponds to the cylindrically symmetric
expansion and is governed by the scaling parameters $a_{x}=a_{y}=a/\sqrt{2}$%
, $a_{z}=b$ . For $a>>b$ we have the case of an initially disc-shaped cloud while for $b\gg a$ we
are effectively in the cigar-shape limit. An isotropic limit obviously
corresponds to $b=a/\sqrt{2}$.

In the anisotropic cylindrically symmetric case Eqs.(\ref{eq.27}) read 
\begin{equation}
\ddot{a}a/2=\ddot{b}b=\frac{\lambda }{\left( a^{2}b/2\right) ^{2/3}}.
\label{eq.44}
\end{equation}%
Note that in the initial moment of expansion when the trapping potential is
switched off%
\[
\frac{b}{a}|_{t=0}=\frac{\omega _{\perp }}{\sqrt{2}\omega _{z}}. 
\]%
The effective potential in accordance with Eq. (\ref{eq.31}) is given by 
\[
U=\frac{3\lambda }{2\left( a^{2}b/2\right) ^{2/3}}. 
\]%
The corresponding Newton equations in agreement with (\ref{eq.44}) acquire
the form%
\[
\ddot{a}=-\frac{\partial U}{\partial a},\,\,\ddot{b}=-\frac{\partial U}{%
\partial b}. 
\]%
Note that this system belongs to the so called Ermakov type of equations 
\cite{Ermakov1880differential}. These equations describe the motion of two
degrees of freedom and therefore to integrate this system it is enough one
to have two autonomous integrals of motion which should be in involution. In
our case, however, we have three integrals of motion. The first one is the
total energy, 
\begin{equation}
E=\frac{1}{2}(\dot{a}^{2}+\dot{b}^{2})+\frac{3\lambda }{2\left(
a^{2}b/2\right) ^{2/3}}.  \label{eq.50}
\end{equation}%
The second and the third integrals are two constants $C_{1},C_{2}$ which
appear in (\ref{eq.13}) while the double integration over time of the virial
identity (\ref{eq.1}),%
\begin{equation}
\frac{d^{2}}{dt^{2}}(a^{2}+b^{2})=4E.  \label{eq.51}
\end{equation}%
The integrals (\ref{C-1and C-2}), however, are not autonomous, they contain
explicit dependence of time, and therefore can not provide a complete
integration of the system. As we will see only their combination defines the
needed integral of motion for the Ermakov type of equations.

Let us introduce now the polar coordinates for $a$ and $b$ 
\[
a=r\cos \Phi ,\text{ }b=r\sin \Phi . 
\]%
In these variables the virial theorem (\ref{eq.51}) acquires the evident form%
\[
\frac{d^{2}}{dt^{2}}r^{2}=4E. 
\]%
where the total energy $E$ in accordance with (\ref{eq.50}) is 
\begin{equation}
E=\frac{1}{2}(\dot{r}^{2}+r^{2}\dot{\Phi}^{2})+\frac{3\lambda }{%
2^{1/3}r^{2}\left( \cos ^{2}\Phi \sin \Phi \right) ^{2/3}}.  \label{eq.54}
\end{equation}%
and correspondingly%
\begin{equation}
r^{2}=2Et^{2}+C_{1}t+C_{2},\,\,C_{1}=\frac{d}{dt}r^{2}-4Et.  \label{eq.55}
\end{equation}%
Multiplying now (\ref{eq.54}) by $r^{2}$ and using the relations (\ref{eq.55}%
) simple calculations give that the combination 
\[
\widetilde{E}=Er^{2}-\frac{1}{2}r^{2}\dot{r}^{2}=EC_{2}-C_{1}^{2}/8 
\]%
is a constant (the Ermakov integral). As the result, we arrive at
conservation law for new "energy" 
\begin{equation}
\widetilde{E}=\frac{1}{2}\left( \frac{d\Phi }{d\tau }\right)
^{2}+U_{eff}(\Phi ),  \label{Ermakov}
\end{equation}%
with new time $\tau $%
\begin{equation}
d\tau =\frac{dt}{r^{2}},\text{ where }\tau =\int_{0}^{t}\frac{dt^{\prime }}{%
2E(t^{\prime })^{2}+C_{1}t^{\prime }+C_{2}}  \label{eq.59}
\end{equation}%
where 
\begin{equation}
U_{eff}(\Phi )=\frac{3\lambda }{2^{1/3}\left( \cos ^{2}\Phi \sin \Phi
\right) ^{2/3}}  \label{eq.61}
\end{equation}%
plays a role of potential energy. It is always positive and goes to infinity
for $\Phi \rightarrow 0$ and $\Phi \rightarrow \pi /2$. \ The minimum of $%
U_{eff}(\Phi )=9\lambda /2$ corresponds to the isotropic case when $\sin
\Phi _{\min }=1/\sqrt{3}$ $.$ Graphically the effective potential $%
U_{eff}(\Phi )$ is shown on Fig.3. 
\begin{figure}[b]
\label{fig1} 
\centerline{
		\includegraphics[width=0.45\textwidth]{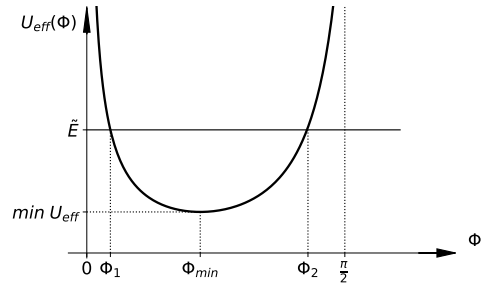}}
\caption{Effective potential $U_{eff}(\Phi )$.}
\end{figure}

The new time $\tau $ \ (\ref{eq.59}) can be easily expressed \ through $t$ ,%
\[
\sqrt{2\widetilde{E}}\tau =\arctan \frac{\sqrt{2E}(t+t_{0})}{\chi }-\arctan 
\frac{\sqrt{2E}t_{0}}{\chi } 
\]%
where $\chi ^{2}=\widetilde{E}/E$ and $t_{0}=\frac{C_{1}}{4E}$ so that $\tau
=0$ at $t=0$. If the initial velocity is equal zero (that is typical for
experiment) the constant $C_{1}=0$ and 
\[
\sqrt{2\widetilde{E}}\tau =\arctan \frac{\sqrt{2\widetilde{E}}t}{C_{2}}. 
\]%
In this case, asymptotically at $t\rightarrow \infty $%
\begin{equation}
\tau \rightarrow \tau _{\infty }=\frac{\pi }{2\sqrt{2\widetilde{E}}}.
\label{eq.64}
\end{equation}%
The trajectory $\Phi (\tau )$ is defined from integration of (\ref{Ermakov})%
\[
\tau =\int \frac{d\Phi }{\sqrt{2\left[ \widetilde{E}-U_{eff}(\Phi )\right] }}%
. 
\]%
Hence the $\tau $-period of the oscillations in the potential $U_{eff}(\Phi
) $ (\ref{eq.61}) is expressed through the integral%
\[
T=2\int_{\Phi ^{(-)}}^{\Phi ^{(+)}}\frac{d\Phi }{\sqrt{2\left[ \widetilde{E}%
-U_{eff}(\Phi )\right] }},  
\]%
where $\Phi ^{(\pm )}$ are roots of equation $\widetilde{E}=U_{eff}(\Phi )$
(reflection points). This integral is expressed via elliptic integrals of
the third order  (see \cite{anisimov1970expansion}). At large value of $%
\widetilde{E}$ oscillations are almost independent on the details of $%
U_{eff}(\Phi )$. Asymptotically in this case the angular velocity $\frac{%
d\Phi }{d\tau }\rightarrow $ $\pm \sqrt{2\widetilde{E}}$ and the $\tau $%
-period 
\[
T\rightarrow \frac{\pi }{\sqrt{2\widetilde{E}}}. 
\]%
namely, in this limit $T$ exceeds in two times $\tau _{\infty }$ (\ref{eq.64}%
). Notice also that dependence of $T$ with respect to $\widetilde{E}$ is
monotonic for the given potential $U_{eff}(\Phi )$ with the maximum
corresponding to the potential minimum. This means that in the
real experiment (what we will discuss in the next Section)  in the better
case it is possible one to observe only half of such oscillation, $t_{osc}$.
Important, that a recurrence to the initial shape is impossible in this
case. In the quasi-classical regime, the gas shape behavior will be different
for cigar and disk initial conditions. For example, in the cigar case we
start at fixed $\widetilde{E}$ from the left reflection point of the
potential $U_{eff}(\Phi )$, in the disk case -- from the right reflection
point. Therefore the shape forms will coincide only for intermediate moments
of time,  far from the initial reflection points.  We should take into
account that at fixed $\widetilde{E}$ starting from any reflection point we
can not reach its opposite reflection point.

$\allowbreak $ $\allowbreak $ $\allowbreak $ $\allowbreak $ $\allowbreak $

It should be emphasized that the solution presented here was obtained first
time by Anisimov and Lysikov \cite{anisimov1970expansion} for expansion of
ideal gas with $\gamma =5/3.$

\subsection{The general anisotropic case}

In the general anisotropic case, when all the scaling parameters are
different $a_{x}\neq a_{y}\neq a_{z}$ it is convenient to introduce the
spherical coordinates ( $r,\theta ,\varphi $) where the total energy
acquires the form%
\begin{eqnarray*}
E &=&\frac{1}{2}\left\lfloor \left( \frac{dr}{dt}\right) ^{2}+r^{2}\left( 
\frac{d\theta }{dt}\right) ^{2}+r^{2}\sin ^{2}\theta \left( \frac{d\varphi }{%
dt}\right) ^{2}\right\rfloor \\
&&+\frac{3\lambda }{2^{1/3}r^{2}}\frac{1}{\left( \sin ^{2}\theta \cos \theta
\sin 2\varphi \right) ^{2/3}}.
\end{eqnarray*}%
Correspondingly introducing again Ermakov reduced energy $\widetilde{E}$ ,
being a sequence of the dilatation symmetry, and new time $\tau $, according
the same prescriptions as in the preceding subsection, we get%
\begin{equation}
\widetilde{E}=C_{2}E-\frac{1}{8}C_{1}^{2}=\left( \frac{d\theta }{dt}\right)
^{2}+\sin ^{2}\theta \left( \frac{d\varphi }{dt}\right) ^{2}+U_{eff}
\label{2degrees}
\end{equation}%
where the effective potential is now%
\begin{equation}
U_{eff}=\frac{3\lambda }{2^{1/3}\left( \sin ^{2}\theta \cos \theta \sin
2\varphi \right) ^{2/3}}.  \label{effective}
\end{equation}%
Thus, we arrive to the system for two degrees of freedom. As it was pointed
out in the previous subsection the integral (\ref{2degrees}) is a
consequence of the scaling symmetry, but for integration of the system it is
not enough. As it was shown by Gaffet \cite{gaffet1996expanding}, this
system indeed has one additional integral (besides $\widetilde{E}$) which
follows from the Painleve test. Existence of these two integrals of motion
guarantees complete integration of this system. As in the previous limit
motion in potential (\ref{effective}) remains its nonlinear
quasi-oscillation character.

\section{Discussion of  experimental data and
comparison with  obtained results}

\begin{figure}[tbh]
\begin{center}
\includegraphics[width=0.25\linewidth]{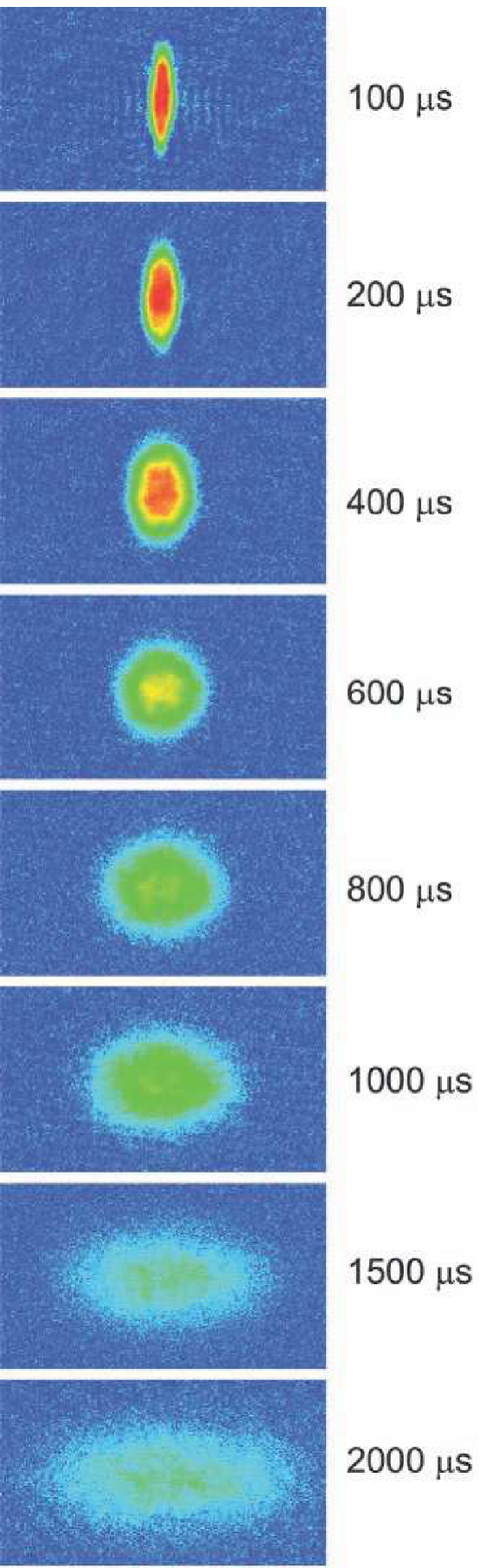} %
\includegraphics[width=0.72\linewidth]{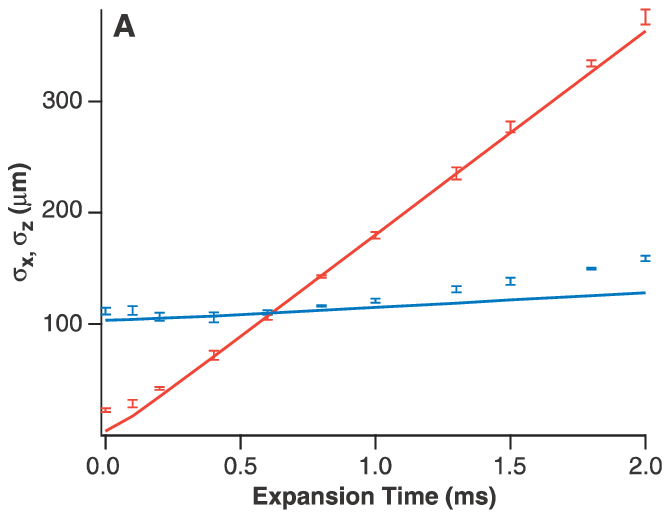}
\end{center}
\caption{(a) Images of a strongly-interacting Fermi gas, which expands
staring from the cigar shape. The expansion time is noted by each image. (b)
The Thomas--Fermi radii along the transverse ($\protect\sigma _{x}$, red)
and longitudinal ($\protect\sigma _{z}$, blue) direction vs the expansion
time. The markers are the data. The curves are the self-similar-expansion
model without adjustable parameters. From~\protect\cite{o2002observation}.}
\label{fig:ThomasExpansion}
\end{figure}

The self-similar expansion of a strongly interacting Fermi gas from a
cigar-shaped trap was observed in \cite{o2002observation}. The images of the
expanding gas are shown in Fig.~\ref{fig:ThomasExpansion}(a). The transverse
size grows rapidly, while the longitudinal size is nearly stationary, with a
weak growth. In Fig.~\ref{fig:ThomasExpansion}(b) one may see qualitative
agreement between the time behavior of the gas expending shape and the
self-similar-expansion model represented by Eqs.~(\ref{eq.17}), (\ref{eq.44}%
). 
The cloud images are changing on Fig.~\ref{fig:ThomasExpansion}(a) from
almost ellipsoid significantly stretched along $z$-axis (exposition $t=100$ $%
\mu s$), later on to the almost spherical shape (at $t=600$ $\mu s$) and,
finally, from the spherical shape to the ellipsoid stretched now in the
direction perpendicular to $z$. The total time of the observation was 2000 $%
\mu s$ which can be taken as a half period (or less) of the period of the
angular shape oscillations, $t\leq t_{osc}/2$, in accordance with the
results of the previous section. The frequency ratio (and thus the
anisotropy ratio up to factor $\sqrt{2}$) in the experiments \cite%
{o2002observation} was rather large initially (around 30) that follows from
the Thomas-Fermi estimation.

Small deviation of the data from the self-similar behavior in Fig.~\ref%
{fig:ThomasExpansion}(b) has been attributed~\cite{zhang2009quantum} to the
contribution of quantum pressure (\ref{eq.7}) into hydrodynamic model (\ref%
{eq.5}), (\ref{eq.6}). That $T_{QP}$ term is neglected to obtain
self-similar solution of Eqs. (\ref{eq.17}), (\ref{eq.44}). This difference,
however, may be explained even without the quantum pressure, by possible
deviation of the equation of state from the $\mu \propto n^{2/3}$
dependence~(\ref{eq.3}) since in experiment the interaction parameter is not
tuned exactly on resonance $1/(k_{F}a_s)=0$, with the estimate $%
1/(k_{F}a_s)\simeq -0.14$ \cite{o2002observation}.

Exactly on resonance, the mean squared cloud size $\langle \mathbf{r}%
^{2}\rangle \equiv \langle x^{2}\rangle +\langle y^{2}\rangle +\langle
z^{2}\rangle $ is found \cite{ThomasScaleInvariance2014} to evolve as 
\begin{equation}
\langle \mathbf{r}^{2}\rangle =\langle \mathbf{r}^{2}\rangle _{t=0}+\frac{%
t^{2}}{m}\langle \mathbf{r}\cdot \nabla U(\mathbf{r})\rangle _{t=0},
\label{eq:OnResTotalSizeExpansion}
\end{equation}%
where $U(\mathbf{r})$ is the initial trapping potential. Expansion law (\ref%
{eq:OnResTotalSizeExpansion}) was obtained within the Thomas-Fermi
approximation and coincides with the quasi-classical dependence for $\langle 
\mathbf{r}^{2}\rangle $ (\ref{quasi-virial}) in the unitarian limit. It
should be emphasized that, according to (\ref{quasi-virial}), $\langle 
\mathbf{r}^{2}\rangle $ indeed depends linearly on energy $E$ that was
verified in experiments \cite{ThomasScaleInvariance2014}.

In the case when the system is far from the unitarian point $%
(k_{F}a_{s})^{-1}=0$ experiments nevertheless give the parabolic time
dependence for $\langle \mathbf{r}^{2}\rangle $. Small deviation of the data
from the self-similar behavior in Fig.~\ref{fig:ThomasExpansion}(b) has been
attributed~\cite{zhang2009quantum} to the contribution of quantum pressure (%
\ref{eq.7}) into hydrodynamic model (\ref{eq.5}), (\ref{eq.16}). That $%
T_{QP} $ term is neglected to obtain self-similar solution of Eqs. (\ref%
{eq.17}), (\ref{eq.44}). This difference, however, may be explained even
without the quantum pressure, by reasons which include deviation of the
equation of state from the $\mu \propto n^{2/3}$ dependence~(\ref{eq.3})
since in experiment the interaction parameter is not tuned exactly on
resonance $1/(k_{F}a)=0$, with the estimate $1/(k_{F}a)\simeq 1/(k_{\text{FI}%
}a)=-0.14$, where $k_{\text{FI}}$ is the Fermi wave vector of a
noninteracting Fermi gas with the same atom number and in the same trap \cite%
{o2002observation}.
\begin{figure}[tbh]
\begin{center}
\includegraphics[width=\linewidth]{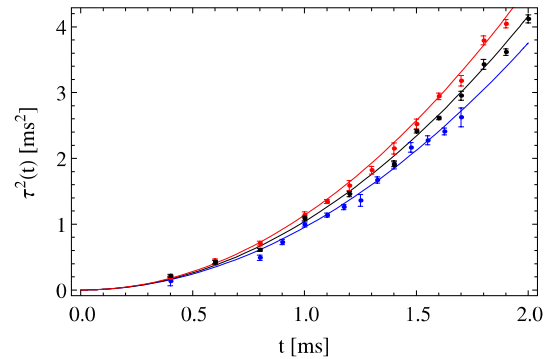}
\end{center}
\caption{The data are experimental values $\protect\tau ^{2}(t)\equiv
m[\langle \mathbf{r}^{2}\rangle -\langle \mathbf{r}^{2}\rangle
_{t=0}]/\langle \mathbf{r}\cdot \protect\nabla U\rangle _{t=0}$ measured for
a strongly-interacting normal Fermi gas after expansion for time $t$,
initially trapped in potential $U(\mathbf{r})$. Black curve is expansion law
(\protect\ref{eq:OnResTotalSizeExpansion}). Black markers correspond to the
gas on resonance, $1/(k_{F}a)=0$, red and blue markers to $1/(p_{F}a)\simeq
1/(k_{\text{FI}}a)=0.59$ and $1/(k_{F}a)\simeq 1/(k_{\text{FI}}a)=-0.61$,
while the solid curves are the results of a calculation without free
parameters \protect\cite{ThomasScaleInvariance2014}. }
\label{fig:ThomasScaleInvBreaking}
\end{figure}

Expansion law (\ref{eq:OnResTotalSizeExpansion}) is indeed the same as for
the ideal gas and coincides with the virial theorem (\ref{eq.1}) . Note that the
expansion law is obtained for both zero- and finite-temperature gas with
equation of state 
\begin{equation}
P=\frac{2}{3}\mathcal{E},  \label{eq:UniversalEOS}
\end{equation}%
where $P$ is the pressure and $\mathcal{E}$ is the energy density, while the
relation $\mu \propto n^{2/3}$ is a particular case of (\ref{eq:UniversalEOS}%
) that corresponds to the isentropic regime for $\gamma =5/3$. Equation of
state (\ref{eq:UniversalEOS}) and expansion law (\ref%
{eq:OnResTotalSizeExpansion}) are consequences of resonant interaction with $%
1/(k_{F}a)=0$. Away from resonance, the expansion laws slightly differ from each other,
which is seen in the measurements displayed in Fig.~\ref%
{fig:ThomasScaleInvBreaking}. Nevertheless, these laws for different
parameters $1/(k_{F}a)$ have the same parabolic dependence on $t$. It should
be emphasized that during the expansion the interaction parameter $1/(k_{F}a_s)
$ changes due to a drop in gas density. When the parameter values fall
outside the interval $(-1.1)$, the quantum effects become less significant,
and the gas expansion approaches the law for a classical monoatomic gas,
which coincides, however, with (\ref{eq:OnResTotalSizeExpansion}).  By these reasons 
we guess that expansions with  $1/(k_{F}a)\simeq
1/(k_{\text{FI}}a_s)=0.59$ and $1/(k_{F}a_s)\simeq 1/(k_{\text{FI}}a_s)=-0.61$  (red and blue curves
in Fig.~\ref{fig:ThomasScaleInvBreaking}) correspond to normal Fermi gas.

\section{Conclusion}

We have demonstrated that symmetry for the GPE in the unitarian limit, describing 
strongly interacting superfluid Fermi gas,
provides existence of the virial theorem ((\ref{eq.1})). As its consequence,  independently on
the ratio between quantum pressure and chemical potential while the
Fermi superfluid gas expansion the size of the gas cloud scales linearly
with time asymptotically so that the expansion velocity tends to the
constant value, $v_{\infty }=\left( 2H/N\right) ^{1/2}$.

For description of the expansion of the
strongly interacting superfluid Fermi gas we have applied the self-similar quasiclassical theory.  
For large time scales the theory
matches quite well with simple ballistic ansatz and also with the
initial quasi-classical distribution of trapping gas. This self-similar solution
is a consequence of the scaling symmetry of the Ermakov type. At large times
the size of the gas clouds scales linearly with time that is a consequence
of the virial theorem. In the unitary limit, when both kinetic and potential
energy scale linearly with the Fermi energy, our quasiclassical solution for
superfluid quantum gas coincides with the Anisimov-Lysikov solution \cite%
{anisimov1970expansion} for classical gas expansion in the isentropic
regime. This anisotropic solution describes the nonlinear deformations of
the cloud shape while self-similar gas expansion.  
For the initial condition in the cigar-shape form this
solution demonstrates successively all the stages of gas expansion, starting
from the distribution extended along the cigar axis, bypassing the
spherically symmetrical one and ending with the distribution, turned at
angle $\pi /2$ with respect to the initial cigar form. Such behavior was
observed first time in experiments \cite{o2002observation}. For the initial
distribution in the form of a quasi-2D disk, all stages of expansion are
inverse to those for the initial distribution in the cigar form.

In order to understand the role of the quantum pressure while the Fermi gas expansion
we would like to note  that the GPE (\ref{eq.2}) admits also the following
self-similar substitution,  
\begin{equation}
\psi =\frac{1}{t^{3/4+i\nu }}F\left( \frac{\mathbf{r}}{\sqrt{t}}\right) ,
\label{similar-anzats}
\end{equation}%
where $\nu $ plays the role of the nonlinear eigenvalue for the differential
equation for the function $F\left( \mathbf{\xi }\right) $ which is assumed
to vanish at large $\mathbf{\xi =r/}\sqrt{t}$. For this substitution the
nonlinear interaction term in (\ref{eq.2})  is of the same order as the
quantum pressure one. However, this ansatz contradicts to the relation (\ref%
{eq.13}) following from the virial theorem. If one substitutes (\ref%
{similar-anzats}) in (\ref{eq.13}) it follows immediately that the quantity $%
\left\langle r^{2}\right\rangle $ grows linearly in time but $\left\langle
r^{2}\right\rangle $ must grow quadratically as $t\rightarrow \infty $
because the Hamiltonian $H$ is strictly positive. This means that (\ref%
{similar-anzats}) can not be applied for description of the system for the
whole space. We may hope only that this ansatz can be used in some region
as, for instance it happens for the cubic NLSE describing weak collapse
regime \cite{zakharov1986quasi}. By this reason, we can suppose that account
of the quantum pressure should provide a transition to the quasi-classical
self-similar asymptotics because in the general case $\left\langle
r^{2}\right\rangle $ behaves similarly in the quasi-classical limit, compare
with (\ref{quasi-virial}). The verification of this assumption will be a
subject of our future numerical study.
\subsection{Acknowledgments}
The work of E.A.K. was supported by the Russian Science Foundation (grant
19-72-30028), M.Yu.K is grateful for the support to the Russian Science
Foundation under grant 18-12-00002, A.~V.~T. was supported by the Presidium of
the Russian Academy of Sciences (Program ``Newest methods of mathematical modeling of nonlinear
dynamic systems'').


\end{document}